\documentclass[conference]{llncs}
\usepackage{cite}
\usepackage{amsmath,amssymb,amsfonts}
\usepackage{algorithmic}
\usepackage{graphicx}
\usepackage{textcomp}
\usepackage{xcolor}
\begin{document}

\title{LLM-based Behaviour Driven Development for Hardware Design}
%\thanks{Parts of this work have been supported by DFG within in Reinhart Koselleck Project {\em PolyVer: Polynomial Verification of Electronic Circuits} (DR 287/36-1).}}

%\author{\IEEEauthorblockN{Rolf Drechsler \hspace*{2cm} Qian Liu}
%\IEEEauthorblockA{\textit{University of Bremen/DFKI, 28359 %Bremen, Germany} }
%drechsler@uni-bremen.de
%}
%\author{}

%\title{Towards LLM-based Generation of Human-Readable Proofs in %Polynomial Formal Verification\\ {\small (Invited Paper)}}
\author{Rolf Drechsler\orcidID{0000-0002-9872-1740}
\and
Qian Liu\orcidID{0009-0000-8071-9039}}%\inst{2,3}\orcidID{1111-2222-3333-4444} \and
%Third Author\inst{3}\orcidID{2222--3333-4444-5555}
%}
\institute{University of Bremen/DFKI, Bremen, Germany \\   
\email{drechsler@uni-bremen.de} \\
\url{http://www.rolfdrechsler.de} 
}
\maketitle

\begin{abstract}
Test and verification are essential activities in hardware and system design, but their complexity grows significantly with increasing system sizes. While {\em Behavior Driven Development} (BDD) has proven effective in software engineering, it is not yet well established in hardware design, and its practical use remains limited. One contributing factor is the manual effort required to derive precise behavioral scenarios from textual specifications. 

Recent advances in {\em Large Language Models} (LLMs) offer new opportunities to automate this step. In this paper, we investigate the use of LLM-based techniques to support BDD in the context of hardware design.
\end{abstract}

%\begin{IEEEkeywords}
%Behaviour Driven Development, LLM, verification, hardware design 
%\end{IEEEkeywords}

\section{Introduction}

Test and verification are central tasks in modern system design. With the continuous increase in system complexity and size, these tasks are becoming even more critical to ensure functional correctness and reliability. In the software domain, {\em Test Driven Development} (TDD) is a well-established methodology that has demonstrated several advantages in improving quality and reducing development effort~\cite{b6,b7,b8}. As a further evolution of this idea, {\em Behavior-Driven Development} (BDD) has been introduced, offering a more structured and behavior-oriented view on specification and testing~\cite{b8,b9,b10,b11,b5,b14}.

Recently, BDD has also gained attention in the hardware domain~\cite{b1,b3} as well as in hardware–software co-design~\cite{b4}. However, the description of the expected behavior is typically performed manually, which is both time-consuming and error-prone. First attempts to automate this process employed classical {\em Natural Language Processing} (NLP) techniques~\cite{b2}, but these approaches remain limited in their expressiveness and accuracy.

In parallel, {\em Large Language Models} (LLMs) have made remarkable progress and are increasingly being explored in the field of {\em Electronic Design Automation} (EDA)~\cite{b12,b13}. Their ability to interpret and generate complex textual descriptions makes them a promising technology to support BDD in hardware design.

In this paper, we apply LLM-based techniques to BDD for hardware. We demonstrate that, starting from a textual specification, an LLM can automatically generate high-level behavioral descriptions suitable for BDD workflows. As a case study, we show for an {\em Arithmetic Logic Unit} (ALU) that relevant scenarios can be generated directly by the LLM and subsequently simulated on the corresponding Verilog implementation.

\section{Scenario Generation}

A central element of BDD is the definition of {\tt scenarios} that describe the expected system behavior in a structured and executable form. In practice, these scenarios are commonly written using the Gherkin language. A typical Gherkin feature file may look as follows:

{\footnotesize
\begin{verbatim}
Feature: User Login

  Scenario: Successful login with valid credentials
    Given the user is on the login page
    When the user enters valid credentials
    Then the user should see the dashboard
\end{verbatim}
}

Such scenarios must traditionally be crafted manually, a process that is both time-consuming and prone to inconsistencies or omissions, as noted in the introduction. Earlier attempts to automate scenario derivation relied on classical AI and NLP techniques~\cite{b2}. However, these methods were limited in their ability to extract meaningful behavioral descriptions from non-trivial specifications and required substantial fine-tuning to achieve acceptable results.

With recent advances in NLP models, the generation of BDD scenarios can be significantly simplified and improved. In our approach, the initial textual specification of the hardware design is provided directly to a hybrid generation system. For this work, we employ ChatGPT-5\footnote{https://openai.com/}
%, Google Gemini\footnote{https://gemini.google.com/}and Groq-based inference\footnote{https://groq.com/} 
together with a lightweight local template engine. The prompting process is intentionally minimal, enabling the system to interpret the specification and produce suitable Gherkin scenarios without extensive engineering effort. A representative prompt is shown below:
\begin{center}
%\emph{Prompt:} ``I want to test the design based on behavior-driven development;  can you specify it accordingly in Cucumber?''

%ql I modified the prompt message replacement.
\emph{Prompt:} ``Create ADD scenario with A = B, 3 examples.''
\end{center}
Using this simple instruction, the LLM is able to translate the textual description into structured BDD scenarios that follow the {\tt Given-When-Then} pattern. These generated scenarios serve as an effective starting point for hardware verification, reducing manual workload while increasing consistency and coverage.

\section{Case study: ALU}

To demonstrate the complete flow of our approach from textual specification to executable verification, we consider a 16-bit ALU. First, only an informal textual description of the intended behavior of the ALU is available. This mirrors common industrial scenarios, where specifications are written in natural language and must be manually translated into hardware and corresponding test scenarios. Using our LLM-based methodology, both the Verilog implementation and the associated BDD scenarios are automatically generated.

\subsection{LLM-Derived Verilog Implementation}

Based solely on the textual specification, the LLM generates a syntactically and semantically correct Verilog model of the ALU. This includes all arithmetic operations, status signals, and combinational logic. For illustration, the generated code for the \texttt{ADD} operation is shown below:

\begin{figure}[h]
\begin{center}
%ql Modified Simulation of a Generated Scenario
\includegraphics[scale=0.70]{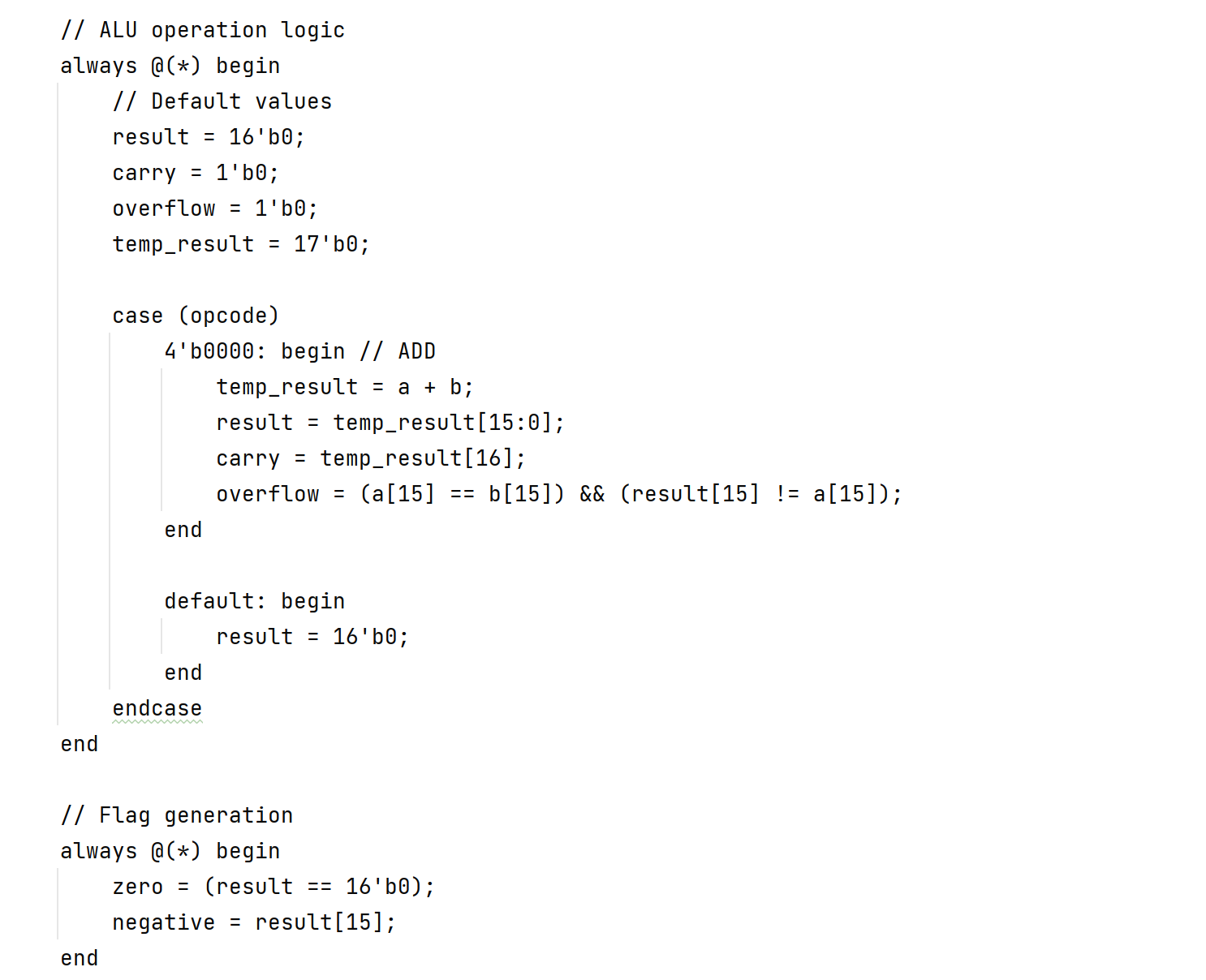}
\end{center}
\end{figure}

The generated Verilog demonstrates several noteworthy properties. First, it adheres to standard design conventions, such as explicit handling of carry propagation and correct two’s complement overflow detection. Second, it reflects structural and stylistic patterns typically used by experienced hardware designers, showing the LLM’s capability to infer and reproduce domain-specific idioms. Finally, we observed that the LLM-generated design was immediately compilable and simulatable without requiring manual debugging, which is an important indicator of practical usefulness.

\subsection{Automatic Scenario Generation}

\begin{figure*}[t]
\begin{center}
% %ql Modified Simulation of a Generated Scenario
\includegraphics[scale=0.60]{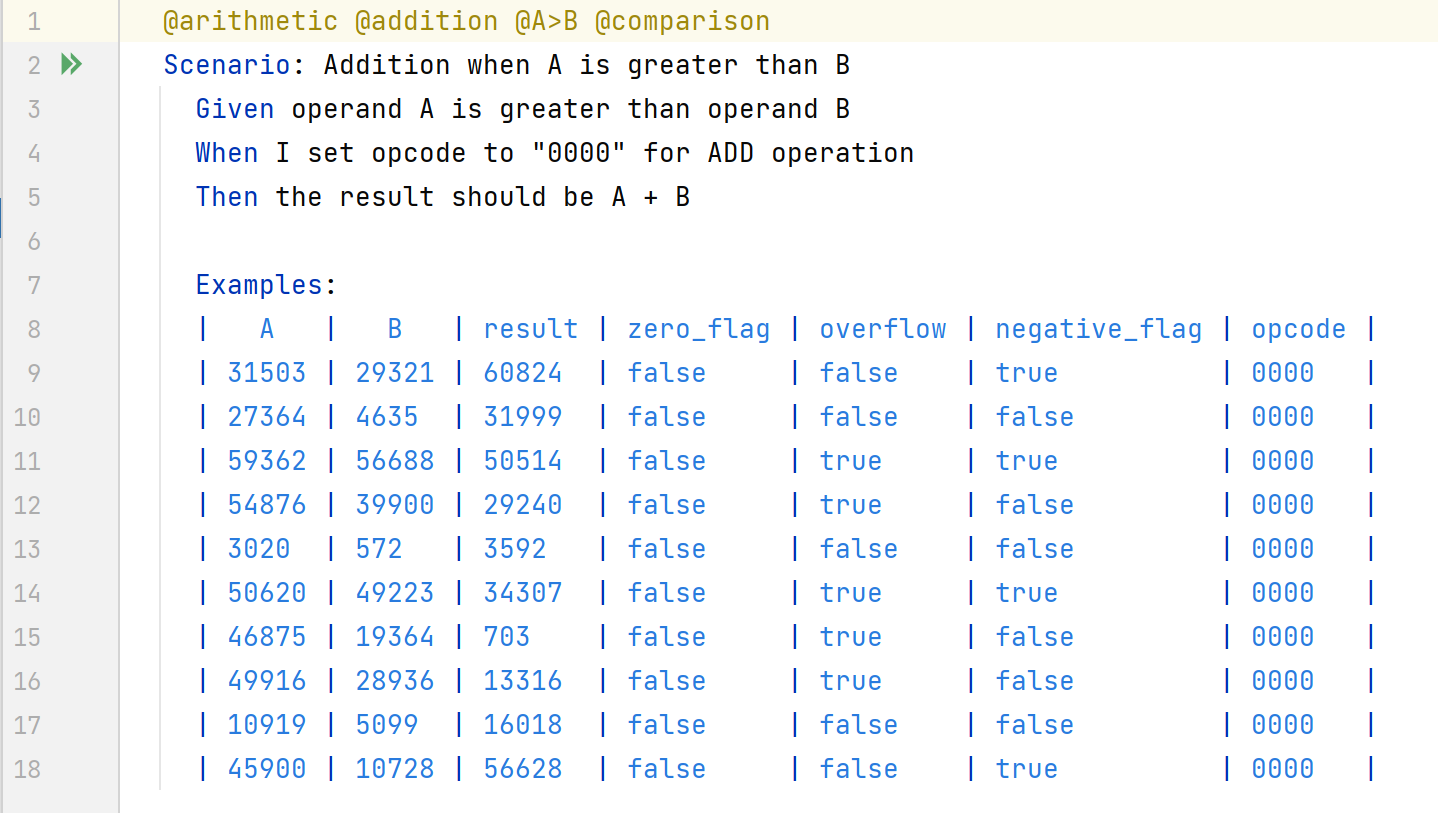}
\caption{Automatically generated scenario for the ADD operation}
\label{fi:scenarioadd}
\end{center}
\end{figure*}

In the next step, BDD scenarios are automatically derived from the same textual specification, ensuring consistency between the implementation and verification intent. These scenarios follow the Gherkin \texttt{Given-When-Then} structure and describe typical as well as boundary-case behaviors. A representative example is shown in Figure \ref{fi:scenarioadd}, where the scenario is described followed by several examples that can be directly used as test cases (see below). 

Similar scenarios are generated for all of the ALU operations.
This illustrates several strengths of the LLM-based approach. First, the model identifies and generates scenarios that test not only the functional outcome but also the status signals that indicate special conditions (e.g., \texttt{carry}, \texttt{zero}, \texttt{overflow}). Second, the generated values span ordinary arithmetic cases as well as edge conditions, such as overflows and subtraction to zero. This indicates that the LLM does not simply restate the specification, but actively reasons about relevant test conditions. Third, the scenarios are expressed in clear Gherkin format, enabling automatic integration into existing BDD and simulation frameworks.

\subsection{Integration with Verilog Simulation}
\begin{figure*}[t]
\begin{center}
% %ql Modified Simulation of a Generated Scenario
\includegraphics[scale=0.35]{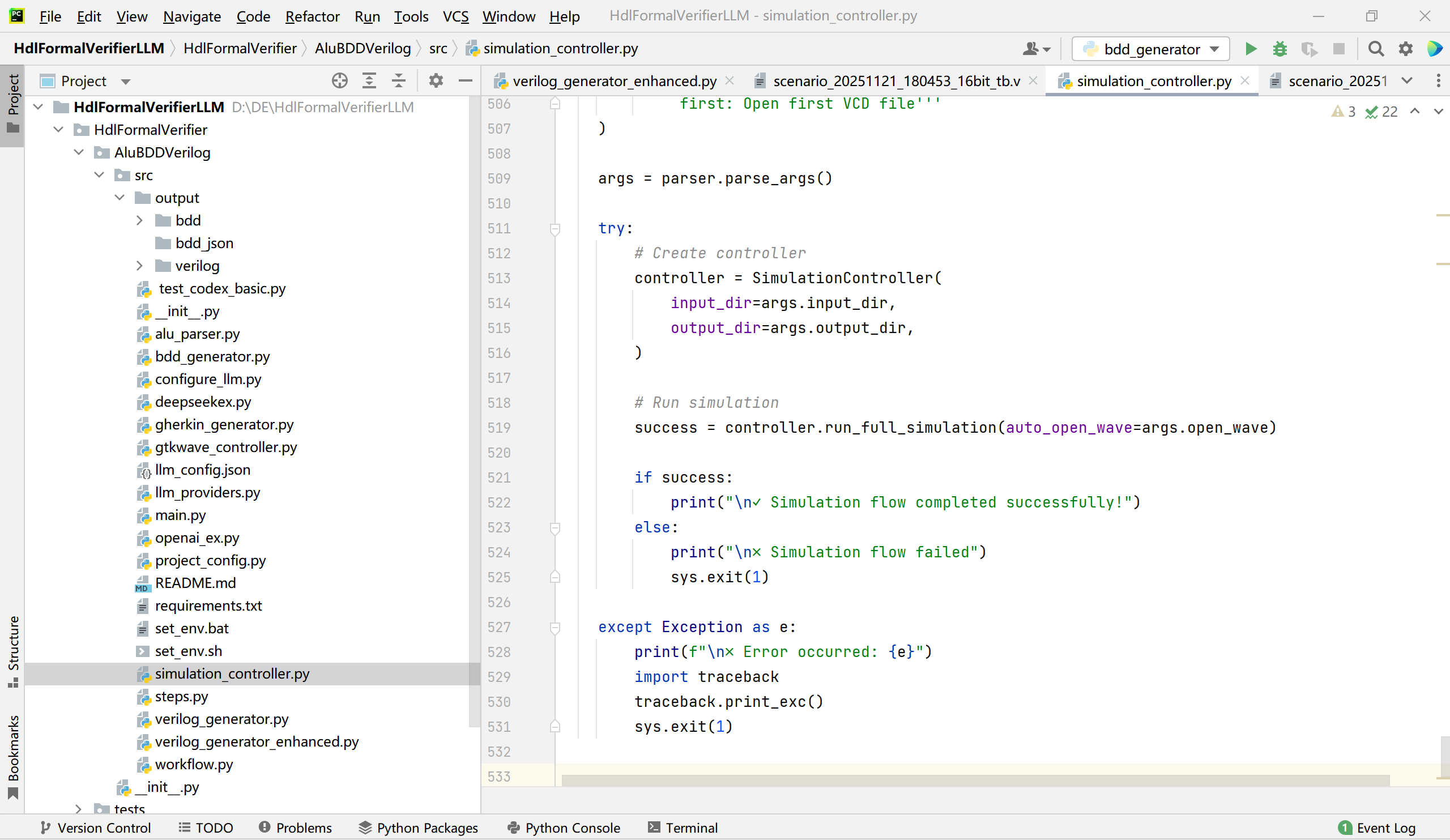}
% %\includegraphics[scale=0.20]{GTKwave.png}
\end{center}
\caption{Development environment}\label{fi:env}
\end{figure*}

\begin{figure*}[t]
\begin{center}
% %ql Modified Simulation of a Generated Scenario
\includegraphics[scale=0.80]{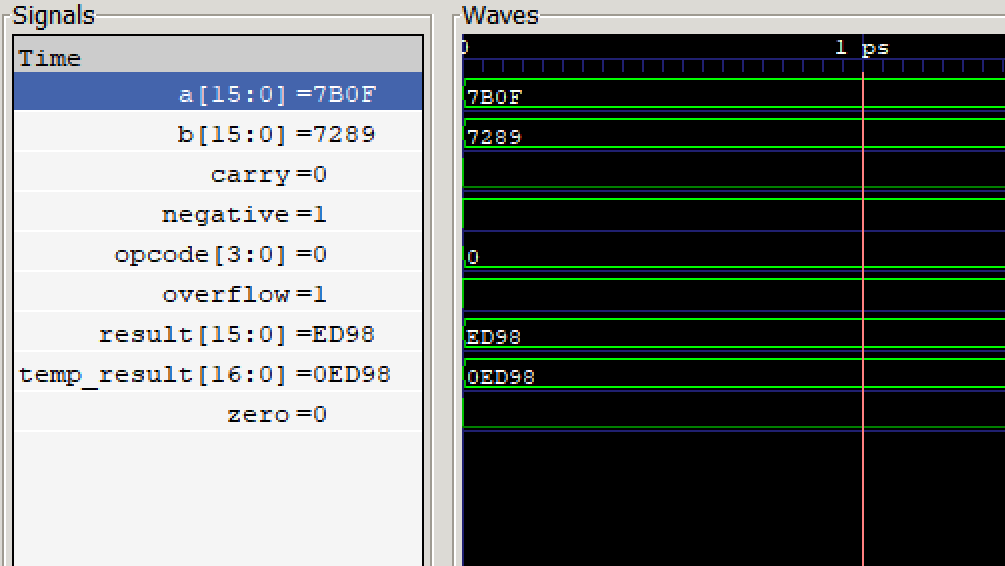}
\end{center}
\caption{Simulation of a generated scenario on the Verilog}
\label{fi:GTKwave}
\end{figure*}

All components have been integrated into a development environment (see Figure \ref{fi:env}). The generated scenarios are then linked to the Verilog implementation using the GTKWave simulator\footnote{https://gtkwave.sourceforge.net/}.Each scenario is translated into stimulus vectors for the ALU, and the resulting waveforms are inspected to verify correctness.Figure~\ref{fi:GTKwave} shows an example waveform corresponding to one of the generated scenarios.

This step demonstrates the practicality of the approach: without manual intervention, the flow proceeds from textual specification to implementation, to test scenario generation, and finally to waveform-level validation. The ability of the LLM to maintain internal consistency across all generated artifacts, i.e.~text, Gherkin, and Verilog, significantly reduces designer workload and lowers the likelihood of human errors that typically arise during specification translation.

\section{Conclusion}

In this work, we have demonstrated that LLMs can effectively support BDD in hardware design. Starting from a purely textual specification, the LLM was able to generate both a synthesizable Verilog implementation and a comprehensive set of BDD scenarios, enabling a seamless flow from specification to simulation. The case study on a 16-bit ALU illustrates the practicality of this approach and highlights the potential to significantly reduce manual effort while improving consistency across design and verification artifacts. These results indicate that LLM-assisted BDD can serve as a promising foundation for more automated and specification-driven hardware development workflows.

While the approach so far is purely based on simulation, based on the BDD scenarios also formal proof techniques can be included by applying symbolic simulations (see e.g.~\cite{b15}).

\section*{Acknowledgment}
This work was supported by the Data Science Center of the University of Bremen (DSC@UB) funded by the State of Bremen.

\end{document}